# The confinement free energies of non-ideal branched polymers and ideal unbranched polymers are the same.

Rouzbeh Ghafouri, Joseph Rudnick, and Robijn Bruinsma

*Department of Physics and Astronomy, University of California, Los Angeles*

*Los Angeles, CA 90095-1547*

## Abstract

We use the method of dimensional reduction to show that a branching polymer with excluded volume interaction confined between two flat plates has, in the thermodynamic limit, a confinement free energy and density profile that is the same as that of an ideal linear polymer with the same number of monomers and the same monomer-plate interaction potential. Condensation due to branching is *exactly compensated* by swelling due to excluded volume interaction.

Branched polymers have been a challenge for statistical mechanics starting from the late 1940's when Zimm and Stockmayer (ZS)[1] showed that the radius of gyration *R(N)* of an ideal branching polymer scales with the number of monomers *N* as $N^{1/4}$. The ZS scaling relation means that ideal branching polymers are *highly condensed*: the monomer density grows linearly with distance from the center of the polymer coil. In turn, that implies that excluded volume interaction should have a very strong swelling effect on branched polymers. A field theory for excluded-volume interactions of branched polymers was constructed[2] in the form of a 6-ε expansion. Using supersymmetry arguments, Parisi and Sourlas[3] (PS) argued that the exponents of a *d* dimensional branched polymer with excluded-volume interaction described by this field theory could be obtained by a mapping to the Yang-Lee edge singularity of a *d*-2 dimensional Ising model. For *d = 3*, this leads to the scaling relation $R(N) \propto N^{1/2}$, with a density profile that now decreases inversely proportional to distance. The supersymmetry method is a demanding formalism but Brydges and Imbrie[4] (BI) showed that it could be reformulated as a relation between the conformational statistics of a branched polymer with excluded volume effects in *d* dimensions to the statistical mechanics of a *hard-core liquid* in *d*-2 dimensions.

The *confinement free energy* of branched polymers plays an important role in the theory of the self-assembly of RNA viruses[5]. In the present paper we use the BI method to compute the confinement free energy and density profile of a branched polymer with excluded volume interaction confined between two plates[6]. The monomers interact with the plates through a potential energy that depends on the distance between the monomers from the plates, e.g., the van der Waals or electrostatic interactions. We will show that the density profile of such a confined branched polymer is the same as that of a confined *ideal* linear polymer having the same number of monomers and the same monomer-plate interaction potential. As a consequence, we find that:
(i) If the monomers interact with the plates via a short-range *repulsive* potential, then the confinement free energy diverges as $1/D^2$ for small *D*, with *D* the plate spacing.
(ii) If the monomers interact with the plates via a short-range *attractive* potential, then there is an *unbinding transition* as a function of the interaction strength[7]. Close to the unbinding transition, the two plates are subject to long-range, polymer-mediated, *bridging attraction*.

(iii) If the interaction potential is a power-law attraction that drops with distance $x$ as $1/x^\alpha$, then there is *always* a surface-adsorbed state if $\alpha < 2$, while for $\alpha > 2$ there is again an unbinding transition.

In order to demonstrate these claims, assume that a freely branching polymer is confined between two plates a distance $D$ apart. The monomers interact with the two plates through an interaction potential $U(x)$, with $x$ the distance from the first plate located at $x=0$. Let $Z_N(x)$ be the number of polymer configurations "rooted" at position $x$, that is, the number of configurations of a branched polymer "grown" from a seed monomer fixed at $x$. $Z_N(x)$ is proportional to the monomer density profile while the integral of $Z_N(x)$ over $x$ is the total partition function. In the BI method, as formulated by Cardy[8], $Z_N(x)$ is related to the coefficients of the fugacity expansion of the density $\rho(x,z)$ of a one-dimensional liquid of hard rods subject to the same potential through:

$$\rho(x,z) = \sum_{N=1}^{\infty} z(-z/\pi)^{N-1} N Z_N(x) \tag{1}$$

It follows from Eq.(1) that, if one knows $\rho(x,z)$, one can reconstruct $Z_N(x)$ from a contour integral surrounding the origin of the complex $z$ plane:

$$Z_N(x) = \frac{(-\pi)^{N-1}}{N} \frac{1}{2\pi i} \oint \frac{\rho(x,z)}{z^{N+1}} dz \tag{2}$$

The contour should be small enough that is does not contain any of the mathematical singularities of $\rho(x,z)$. Since a *physical* $d=1$ liquid of hard rods cannot undergo a phase transition, there can be no singularity along the positive $z$-axis. However, mathematical singularities elsewhere in the complex plane are possible, as we shall see.

In order to determine the analytical structure of $\rho(x,z)$, it is convenient to confine the rods to a *discrete lattice* $n = 0, 1, 2,..., L$ and then later take the continuum limit. We will restrict ourselves to the simplest non-trivial case in which each rod blocks two sites. Local thermodynamic quantities of $d = 1$ systems in general obey *recursion relations*. Recursion relations for a liquid of hard rods in an external potential were constructed by Percus[9,] and modified for computational purposes by Vanderlick et al[10]. For a discrete lattice, these recursion relations adopt the form[11]:

$$h_n = \frac{z\exp^{-\beta U_n}(1-h_{n-1})}{1+z\exp^{-\beta U_n}(1-h_{n-1})} \qquad (3)$$

$$h_n = \frac{p_n}{1-p_{n+1}} \qquad (4)$$

where $p_n(z)$ is the probability that a rod blocks sites $n$-1 and $n$. The first recursion relation has the form of the *Langmuir Adsorption Isotherm*, apart from the $(1-h_{n-1})$ factors that correct the Boltzmann factor for the fact that a rod located at site $n$ blocks site $n$-1. Note that $h_n \simeq 1$ and $p_n \simeq 1/2$ in the limit of large $z$, which corresponds to full coverage. For the *uniform* lattice liquid, with $U_n = 0$, the analytical structure of the occupation probability $p(z)$ in the complex plane is easily obtained by eliminating $h$ from Eqs. (3) and (4):

$$\frac{1-2p(z)}{1-p(z)} = \frac{1}{2z}\left(\sqrt{1+4z}-1\right) \qquad (5)$$

According to Eq.(5), $p(z)$ is analytic around $z = 0$ - as required by Eq.(1) - while it has a branch-cut along the negative real axis starting at $z = -1/4$. The partition function can be evaluated by deforming the contour in Eq. (2) to one that follows the upper and lower bounds of the branch-cut. The resulting integral along the negative z-axis is straightforward, leading to a partition function $Z_N \propto \dfrac{(4\pi)^N}{N^{3/2}}$ consistent with PS scaling exponents.

For a *non-uniform* hard-rod liquid, Eqs. (3) and (4) are less convenient because of their non-linearity. Define a new set of quantities $\phi_n$:

$$1 + z_n(1 - h_n) = (-z_{n+1})^{1/2} \frac{\phi_{n+1}}{\phi_n} \tag{6}$$

where $z_n = z \exp{-\beta U_n}$, and use Eq.(6) to eliminate the $h_n$ from Eq.(3). One obtains a *linear* two-step recursion relation:

$$(-z_{n+1})^{1/2} \phi_{n+1} + (-z_n)^{1/2} \phi_{n-1} - \phi_n = 0 \tag{7}$$

The site occupation probabilities can be expressed in terms of the $\phi_n$ by combining Eqs. (4) and (6):

$$p_n = -\left(\frac{z_{n-1}}{z_n}\right)^{1/2} \frac{\phi_{n-2}}{\phi_n}(1 - p_{n+1}) \tag{8}$$

We impose the boundary conditions $p_0 = p_{L+1} = 0$. It follows from Eq.(6) that the condition $p_0 = h_0 = 0$ is obeyed if $\dfrac{\phi_1}{\phi_0} = (1 + z_0)(-z_1)^{-1/2}$.

In order to determine the analytical structure of $p_n(z)$ in the complex plane, it is useful to carry out the first few steps of Eq.(8), from right to left, starting at $n = L + 1$, and imposing the second boundary condition $p_{L+1} = 0$:

$$p_{L-3} = -\left(\frac{z_{L-4}}{z_{L-3}}\right)^{1/2}\frac{\phi_{L-5}}{\phi_{L-3}} - \left(\frac{z_{L-4}}{z_{L-2}}\right)^{1/2}\frac{\phi_{L-5}\phi_{L-4}}{\phi_{L-3}\phi_{L-2}} - \left(\frac{z_{L-4}}{z_{L-1}}\right)^{1/2}\frac{\phi_{L-4}\phi_{L-5}}{\phi_{L-2}\phi_{L-1}} - \left(\frac{z_{L-4}}{z_L}\right)^{1/2}\frac{\phi_{L-4}\phi_{L-5}}{\phi_{L-1}\phi_L}$$
(9)

If the functions $\phi_n(z)$ are analytic, then the only possible singularities of $p_n(z)$ are *poles* at $z$ values where one of the $\phi_n$ vanish (note that ratios of the $z_n$ are independent of $z$). However, if the last two terms of Eq. (9) are combined into a single term one finds that this term is *proportional* to $\phi_{L-1}$ if one uses the recursion relation Eq.(7), so there is actually no pole associated with the vanishing of $\phi_{L-1}$. Similarly, the second and third term can be combined to a single term that is proportional to $\phi_{L-2}$. This argument can be iterated with the result that $p_n(z)$ is in fact regular when *any* of the $\phi_{L-m}$ vanishes *with the exception of m=0*. The reason for the exception is that the last term Eq. (9) has no pairing partner. It follows that singularities of $p_n(z)$ in the complex plane are restricted to poles located at $z_m$ that are solutions of $\phi_L(z = z_m) = 0$.

In order to locate the solutions of $\phi_L(z = z_m) = 0$, we will focus on complex $z$ values near the termination point $z = -1/4$ of the branch-cut of the uniform system. If $z = -\frac{1}{4}(1+\varepsilon)$ with |ε| << 1 and if the monomer interaction energy is small compared to the thermal energy then $z_n \simeq -\frac{1}{4}(1+\varepsilon - \beta U_n)$, and we can write Eq.(7) as:

$$-\phi_{n+1} - \phi_{n-1} + 2\phi_n + \beta U_n \phi_n = \varepsilon \phi_n \qquad (10)$$

where $\frac{\phi_1}{\phi_0} \simeq (1+z)(-z)^{-1/2} \simeq 3/2$. The analytical form of $p_n(z)$ can now be expressed as:

$$p_n(z) \simeq F_n(z) - \frac{\phi_{n-1}(z)\phi_{n-2}(z)}{\phi_{L-1}(z)\phi_L(z)} \qquad (11)$$

where $F_n(z)$ is a purely analytic function. The condition $\phi_L(z=z_m)=0$ for the location $z=z_m$ of a pole is obeyed if $\phi_n(z_m)$ is a solution of the difference equation Eq.(10) with boundary conditions $\phi_L = 0$ and $\frac{\phi_1}{\phi_0} \simeq 3/2$. The poles of $p_n(z)$ thus correspond to the *eigenvalues* of Eq.(10) for these boundary conditions, while the $\phi_n(z_m)$ are the corresponding *eigenfunctions*.

In the continuum limit, we can replace $n$ with $x/a$, where $a$ is a short distance cutoff - corresponding to the monomer size with $L = D/a$ and where $p_n(z)$ is replaced by $a\rho(x,z)$. The difference equation (10) turns into the *Schrödinger Equation*:

$$-a^2 \frac{d^2}{dx^2}\phi(x,\varepsilon) + \beta U(x)\phi(x,\varepsilon) = \varepsilon \phi(x,\varepsilon) \tag{12}$$

In the continuum limit, the boundary condition at x=0 is $\phi(0,\varepsilon)=0$ [12]. Finally, in Eq.(11) we must replace[13]

$$\phi_{L-1}\phi_L \sim (\varepsilon - \varepsilon_m)\int_0^D \phi(x,\varepsilon_m)^2 \, dx \tag{13}$$

when $\varepsilon$ is close to one of the eigenvalues $\varepsilon_m$ of the Schrodinger Equation. According to Eqs. (11) and (13), $\rho(x,z)$ has a *simple pole* at $z_m = -\frac{1}{4}(1+\varepsilon_m)$. The residue of the pole is equal to $\frac{1}{4}\tilde{\phi}(x,\varepsilon_m)^2$ where $\tilde{\phi}(x,\varepsilon_m)$ is the normalized eigenfunction of the eigenvalue $\varepsilon_m$.

Having established the analytical structure of $\rho(x,z)$, we can now apply the *Mittag-Leffler Theorem*[14], according to which functions that are analytical except for a countable infinite number of poles ("meromorphic functions") can be written as the sum of a purely analytical function plus the principal parts of the pole singularities. For the present case, the theorem implies that $\rho(x,z)$ must have the analytical form

$$\rho(x,z) = F(x,z) - \frac{1}{4}\sum_m \frac{\tilde{\phi}(x,\varepsilon_m)^2}{z - z_m} \tag{14}$$

with *F(x,z)* some analytic function of *z*. Conversely, it can be shown that if one *assumes* an analytical structure for the density of the form of Eq.(14) then the $\tilde{\phi}(x,\varepsilon_m)$ must be normalized eigenfunctions of the Schrödinger equation with eigenvalue $\varepsilon_m$. We now can perform the contour integral Eq.(2) to reconstruct the rooted partition function. This done by first deforming the contour to run again along the two sides of the negative z-axis and then breaking up the integral into a sum of separate contours surrounding each of the poles along the negative z-axis. Summing over the poles, one obtains:

$$\boxed{Z_N(x) \propto \frac{(4\pi)^N}{N} \sum_m \tilde{\phi}(x,\varepsilon_m)^2 \exp^{-N\varepsilon_m}} \tag{15}$$

Note that the unknown analytical contribution *F(x,z)* does not contribute.

By integrating $Z_N(x)$ over *x* from 0 to *D* we obtain the total partition function and, using the normalization condition, we can express the confinement free energy - the *D*-dependent part of the free energy - as $-\ln\left(\sum_m \exp^{-N\varepsilon_m}\right)$. In the thermodynamic limit of large *N*, the sum is dominated by the lowest eigenvalue of the Schrödinger Equation, i.e., the ground-state energy. The confinement free energy reduces to the groundstate energy times *N*.

The claims made in the introduction follow from Eq. (15) straightforwardly by computing the eigenvalues and eigenfunctions for the appropriate Schrödinger Equation. If, for example, the potential is zero then the eigenvalue spectrum is $\varepsilon_m = \pi^2 m^2 a^2 / D^2$ with eigenfunctions $\tilde{\phi}(x,\varepsilon_m) \propto D^{-1/2} \sin(\pi mx/D)$. It then immediately follows that the confinement energy diverges as $1/D^2$ in the limit $D \ll aN^{1/2}$ where the groundstate dominates. In the opposite limit $D \gg aN^{1/2}$, the poles fuse into the branch cut of Eq.(5) and the integral of $Z_N(x)$ over $x$ reduces to $Z_N \propto \frac{(4\pi)^N}{N^{3/2}}$. Next, if for $D \gg aN^{1/2}$, the potential energy $U(x)$ of a single plate is such that the Schrödinger Equation has a *bound state* then the branched polymer will be adsorbed on either of the two plate surfaces. The plate-plate interaction is *attractive* in this regime, the "bridging attraction". This is due to "quantum tunneling" between the two surface bound states, which reduces the eigenvalue and thus the free energy. If the depth of the potential wells is reduced, then the bound states disappear. The bound state pole fuses with the branch cut while the plate-plate interaction becomes repulsive.

It is well known that the conformational statistics of an ideal, *unbranched* polymer in an external potential also can be obtained from solutions of the Schrödinger Equation[15]. Specifically, the number of configurations of an N-monomer ideal polymer placed between the same two plates with the initial monomer located at a distance $x$ and the final monomer at a distance $x'$ from the x=0 plate is given by

$$G_N^{ideal}(x,x') \propto \sum_m \tilde{\phi}(x,\varepsilon_m)\tilde{\phi}(x',\varepsilon_m)\exp^{-N\varepsilon_m} \qquad (16)$$

where coordinates in the plane of the plates have been averaged over. The eigenfunctions and eigenvalues are defined in the same way as before. For $x = x'$, Eq.(16) describes the statistics of a *buckle* or *loop* whose endpoints are confined to a plane at a distance $x$. Note that $G_N^{ideal}(x,x)$ has the same form as our Eq.(15): *the number of branched polymer conformations with excluded volume interactions rooted at a distance x from the plates is proportional to the number of ideal, unbranched chain conformations that start and end at a distance x from the plate*. In the large $N$ limit, where one can assume ground-state dominance, $G_N^{ideal}(x,x)$ is proportional to the density profile of the unbranched idela polymer. In the thermodynamic limit the density profiles of the two systems are thus the same. In fact, the results claimed in the introduction are familiar from studies of the surface adsorption of ideal chains[16]. Physically, this mapping can be understood as a *precise cancellation between the condensation of ideal branched polymers and the swelling induced by excluded volume repulsion*. This compensation already was foreshadowed by the fact that the PS scaling relation for the radius of gyration of a branched polymer coincides with the scaling relation of an ideal linear chain but it is surprising to see that in the large $N$ limit, the confinement free energies and density profiles of the two systems coincide as well.

Acknowledgements. We would like to thank the NSF for support under DMR Grant 04-04507. RG would like to thank John Cardy for helpful discussions.